\documentclass[12pt,preprint]{aastex}
\usepackage{natbib,epsfig}

\begin{document}

\newcommand{\hii}{{\rm H}{\sc ii}}
\newcommand{\uchii}{{\rm UCH}{\sc ii}}
\newcommand{\etal}{{\it et. al.}}
\newcommand{\htwo}{${\rm H_2}$}
\newcommand{\nhone}{NH$_3$(1,1)}
\newcommand{\nhtwo}{NH$_3$(2,2)}
\newcommand{\nhthree}{NH$_3$(3,3)}
\newcommand{\ammonia}{NH$_3$}
\newcommand{\methanol}{CH$_3$OH}
\newcommand{\htwoo}{H$_2$O}
\newcommand{\hsixalpha}{H66$\alpha$}
\newcommand{\thcoone}{$^{13}$CO ${(J=1\rightarrow0)}$}
\newcommand{\ceioone}{C$^{18}$O ${(J=1\rightarrow0)}$}
\newcommand{\ceio}{C$^{18}$O}
\newcommand{\cseo}{C$^{17}$O}
\newcommand{\nh}{NH$_3$}
\newcommand{\um}{$\mu$m}
\newcommand{\percc}{cm$^{-3}$}
\newcommand{\persqcm}{cm$^{-2}$}
\newcommand{\emunits}{pc~cm$^{-6}$}
\newcommand{\kms}{km~s$^{-1}$}
\newcommand{\kmspc}{km~s$^{-1}$~pc$^{-1}$}
\newcommand{\vlsr}{${\rm v}_{\rm lsr}$}
\newcommand{\msun}{M$_\odot$}
\newcommand{\lsun}{L$_\odot$}
\newcommand{\jyperbeam}{Jy beam$^{-1}$}
\newcommand{\mjyperbeam}{mJy beam$^{-1}$}
\newcommand{\jyperbeamkms}{Jy beam$^{-1}$km~s$^{-1}$}
\newcommand{\arcseconds}{$''$}
\newcommand{\pv}{P-V}

\title{Spherical Infall in G10.6-0.4: Accretion Through an Ultracompact \hii\ Region}

\author{Peter K. Sollins\altaffilmark{1}, Qizhou Zhang\altaffilmark{1}, Eric Keto\altaffilmark{1}, Paul T. P Ho\altaffilmark{1}}
\altaffiltext{1}{Harvard-Smithsonian Center for Astrophysics, 60 Garden Street, Cambridge, MA, 02138, psollins@cfa.harvard.edu}

\begin{abstract}

We present high resolution ($0.''12 \times 0.''079$) observations of
the ultracompact \hii\ region G10.6-0.4 in 23~GHz radio continuum and
the \nhthree\ line. Our data show that the infall in the molecular
material is largely spherical, and does not flatten into a molecular
disk at radii as small as 0.03~pc. The spherical infall in the
molecular gas matches in location and velocity the infall seen in the
ionized gas. We use a non-detection to place a stringent upper limit
on the mass of an expanding molecular shell associated with pressure
driven expansion of the \hii\ region. These data support a scenario in
which the molecular accretion flow passes through an ionization front
and becomes an ionized accretion flow onto one or more main sequence
stars, not the classical pressure-driven expansion scenario. In the
continuum emission we see evidence for externally ionized clumps of
molecular gas, and cavities evacuated by an outflow from the central
source.

\end{abstract}


\section{Introduction} \label{sec:intro}

The ultracompact (UC) \hii\ region G10.6-0.4, at a distance of 6.0~kpc
\citep{dow80}, is the brightest member of a complex of variously
compact \hii\ regions \citep{woo89b} in an area of active star
formation. The associated IRAS point source, IRAS 18075-1956, has a
luminosity of $9.2 \times 10^5$~\lsun\ \citep{cas86} and has colors
that meet the criteria of \citet{woo89a} for an \uchii\ region. G10.6
is known to be embedded in a hot molecular core (HMC)
\citep{bra83,ho86,plu92}. The core is thought to contain at least
1200~\msun\ of gas within a radius of 0.2~pc, based on an analysis of
a variety of dust continuum measurements \citep{mue02}, and at least
3300~\msun\ of gas within 1.1~pc based on \ceio\ and \cseo\
measurements \citep{hof00}. Previous studies of the inversion lines of
\ammonia\ have determined that the molecular core is rotating and
collapsing inward toward the \uchii\ region
\citep{ho86,ket87a,ket88,ket90}. In these studies, using the \nhone\
and \nhthree\ lines, rotation is seen at size scales from 1~pc down to
0.08~pc, and infall is detected in the form of red-shifted absorption
seen against the continuum source. \methanol\ and \htwoo\ masers are
seen distributed linearly in plane of the rotation
\citep{wal98,hof96}, while OH masers seem to be distributed along the
axis of rotation \citep{arg00}. In \ceioone, \citet{ho94} see
$10^3$~\msun\ of dense (n$\sim10^6$\percc), rotating gas in a
flattened (0.3$\times$0.1~pc) disk-like structure. At the highest
resolution achieved in earlier work, infall and rotation in the
molecular gas were seen simultaneously in absorption, showing that the
molecular gas was spiraling inward on size scales comparable to the
size of the \uchii\ region. Our new observations resolve the \uchii\
region and show clearly that the molecular gas is infalling mostly
spherically toward the \uchii\ region, with only slow rotation and
little flattening in the plane of rotation.

Recent observations of the ionized gas within the \uchii\ region
showed that the ionized gas is also spiraling inward toward the stars
at the center of the \uchii\ region \citep{ket02a}. Subsequent
theoretical work showed that in small \hii\ regions, the gravitational
effect of the central star(s) can overcome the thermal pressure of the
ionized gas causing the molecular accretion flow to pass through the
\hii\ region boundary and continue inward as an ionized accretion flow
\citep{ket02b}. In this model, the \hii\ region boundary exists as a
standing R-type ionization front within a continuous accretion
flow. These results differ from classical treatments of the pressure
driven expansion of \hii\ regions, which predict outward motion of the
ionized gas as soon as the \hii\ region is formed
\citep{str39,spitzer}. In the classical model for pressure driven
expansion, the \hii\ region boundary, after a very short phase as a
moving R-type front, will develop a characteristic double front
structure composed of an isothermal shock followed by a moving D-type
ionization front. In this model, as the \hii\ region expands, most of
the displaced molecular material remains between the shock and the
ionization front as a dense outward moving shell, which snow-plows
ahead of the \hii\ region. In the alternative model of \citet{ket02b},
however, the accretion flow passes through a standing R-type
ionization front at the \hii\ region boundary and continues toward the
star(s) as an ionized flow. In that case there will be no swept-up,
dense molecular layer at the boundary, and all the molecular gas will
be moving inward. Ionized accretion flows represent a fundamentally
different mode of accretion than that seen in low-mass star
formation. In the ionized accretion flow scenario a main sequence star
accretes gas which was once part of a molecular accretion flow, but
has been photo-ionized as it approaches the star. G10.6 contains the
first example of evidence for a massive accretion flow continuing
through an ionization front. Our new observations constitute a
stringent non-detection of any outward moving molecular layer thus
contradicting the pressure driven expansion model (see Section
\ref{sec:nature}).

\section{Observations} \label{sec:obs}

We observed the \uchii\ region G10.6 with the NRAO Very Large Array
(VLA)\footnote{The National Radio Astronomy Observatory is a facility
of the National Science Foundation operated under cooperative
agreement by Associated Universities, Inc.} on February 1, 2002, with
the phase center at $\alpha(2000)=\rm{18^h10^m28^s.68}, \,
\delta(2000)=-19^o 55'49''.07$. We observed the (3,3) inversion line
of \ammonia\ at 23.870130 GHz with 63 spectral channels of width
48.828 kHz (0.61 \kms) for a total bandwidth of 3.125 MHz (38.7 \kms)
centered on \vlsr$=10$ \kms, and 1.3~cm continuum with a bandwidth of
15.6 MHz. The array was in the A-configuration, yielding a
uniform-weighted synthesized beam of width $0.''12 \times 0.''072$ for
a physical resolution of 0.0034$\times$0.0021~pc or 700$\times$430~AU.

We observed the quasars 3C286, 3C273 and 1733-130 for flux, bandpass
and phase calibration respectively. Self-calibration of the source
amplitudes and phases resulted in a noise level of 0.18 (0.14)
\mjyperbeam\ in the uniform (natural) weighted continuum map, and 1.9
(1.5) \mjyperbeam\ in each uniform (natural) weighted channel map,
about 3 times the thermal noise limit. Expressed as a temperature, our
sensitivity in a natural weighted continuum map is about 25~K and in a
natural weighted channel map is about 280~K. The physical temperature
of the molecular gas around G10.6 is estimated to be only 110~K at the
ionization front \citep{ket90}. Thus, none of the molecular gas is
detectable in thermal line emission in channel maps at the $3\sigma$
level. But since the continuum has a peak brightness temperature of
6900~K, the line absorption can be detected at roughly $25\sigma$ in a
single channel. The quality of the self-calibration solutions and
improvements in the K-band receiver system at the VLA have resulted in
25 times better sensitivity in our \nhthree\ channel maps than in the
previous best existing \nhthree\ data for this source \citep{ket88},
with 3 times better spatial resolution and 2 times better velocity
resolution.

\section{Results} \label{sec:results}

\subsection{The Structure of the \hii\ Region}  \label{sec:cont}

\begin{figure}
\epsscale{1.0}
\plotone{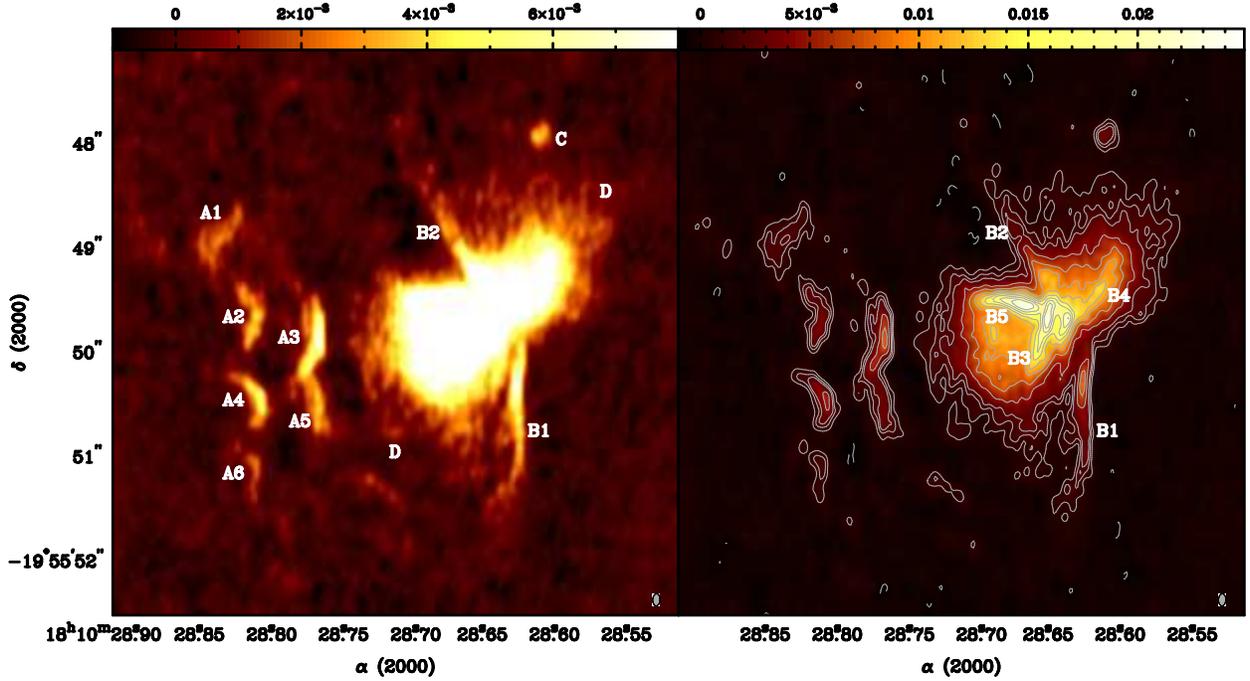}
\caption{The 1.3 cm continuum map. In the left panel, the colors range
linearly from -1 to 8 \mjyperbeam\ although the peak in the map is 24
\mjyperbeam. This is to emphasize the weaker emission structures which
are labeled A1-A6, B1-B5, C and D in the map. In the right panel, the
colors range linearly from -1 to 24 \mjyperbeam, showing the entire
range of emission strengths in order to emphasize structures in the
stronger emission. The contours are -1, 1, 2, 3, 6, 9, 12, 15, 18, 21,
24, \& 27 $\times$ 0.8 \mjyperbeam. The map was made with uniform
weighting and the synthesized beam ($0.''12 \times 0.''079$) is shown
in the lower right of each panel.}
\label{fig:contmap}
\end{figure}

Figure \ref{fig:contmap} shows the 23~GHz continuum map in both
panels. The color-scale in the left panel is stretched to emphasize
the structure in the weaker emission while the right panel is scaled
to show clearly the structure in the strongest emission. Four features
in particular, marked A1-A6, B1-B5, C, and D in Figure
\ref{fig:contmap}, are worth noting.

First, to the east of the main \uchii\ region, there are six arc
shaped continuum sources marked A1 through A6 in Figure
\ref{fig:contmap}. Several of these are arcuate, and convex, pointing
back toward the central \uchii\ region. All except for A1 are
brightest on their western edges as if they are externally ionized by
photons from the \hii\ region or mechanically by a wind. Arcs A2-A5
show line absorption (see Figure \ref{fig:mom1}) suggesting that they
are embedded inside the HMC. Because A1 is also bright enough to show
detectable absorption, but does not, A1 must either be in front of the
densest gas, or the gas toward A1 might have lower column density or
might not be warm enough to populate the (3,3) state. The upper limit
on the optical depth of the \nhthree\ main hyperfine line toward A1 is
$\tau_\nu < 0.6$. By contrast, toward the continuum peak, we detect an
optical depth of the main hyperfine line in excess of 65.

Second, there are larger scale linear structures extending to the
south and northeast from the main \uchii\ region marked B1 and B2 in
Figure \ref{fig:contmap}. In the right panel of Figure
\ref{fig:contmap} there are smaller scale linear features marked B3,
B4 and B5. One can envision a wide angle outflow with an evacuated
cavity whose edges are defined by B1, B3 and B4 on the southwest and
by B2 and B5 on the northeast, parallel to the axis of rotation of the
molecular gas, which is about 40 degrees east of north
\citep{ket87b,ket88}. The outflow would fill the ``V'' shape on the
northeast edge of the \uchii\ region. There is no \nhthree\ absorption
above the sensitivity limit of optical depth 0.3 toward source B1. We
suggest that B1 defines the edge of the cavity evacuated by the
southwest side of the outflow, protruding toward the observer out of
the densest molecular gas. Ongoing observations in the \hsixalpha\
line in this region have measured the motion of ionized gas in the
cavities, with the southwest side (B1, B3, B4) moving toward the
observer, and the northwest side (B2, B5) moving away. The general
shape of the brightest continuum emission is extended perpendicular to
the rotation axis with an aspect ratio of about 2, which is consistent
with there being an ionized disk perpendicular to the proposed outflow
along the rotation axis.

Third, the slightly resolved, almost circular peak just to the
northwest of the main \hii\ region, marked C in Figure
\ref{fig:contmap}, may be a separate ``hyper-compact'' \hii\
region. Because of its regular shape, it seems unlikely that the
ionization is caused by UV photons leaking out of the central \hii\
region. If this is a distinct, extremely compact \hii\ region, its
radius, $\sim 0''.08$ or about 500~AU, is consistent with
gravitational trapping of the ionized gas by 100~\msun\ of stars
\citep{ket02b}.

Fourth, while there are some sharp edges in the continuum emission,
for instance, in sources B1 and B2, the edges of the main \uchii\
region itself are quite gradual, even ``frothy''. This froth is marked
D in Figure \ref{fig:contmap}. This frothy edge structure is present
all around the \uchii\ region except in the regions which we propose
to be the evacuated outflow cavity. So in a scheme in which accretion
is going on through a flattened structure and outflow is going on
along the rotation axis, we can associate the froth with
irregularities specifically in that part of the ionization front
through which accretion is taking place. The irregular structures
could be related to varying density in the gas crossing the ionization
front, or photo-ionization of a very clumpy circumstellar
environment. Several possible ionization front instabilities are
discussed in \citet{ket89}, but because the continuum in this part of
the map is weak we do not have line data to help us distinguish
between different possibilities.

The much sharper ionization fronts associated with B1 and B2 could be
due to the much higher velocities in the outflow and an evacuated
environment in the direction of the outflow. Finally the isolated
nature of A2-A6, with no intervening ionized structures to the central
source, also suggests a clumpy circumstellar environment. To be
consistent with the frothy structure seen towards the central region,
the size-scale and separation of clumps must increase away from the
center. Future work, in which we will map the radio continuum at high
resolution in a variety of frequencies, will allow us to carefully
determine the spectral energy distribution for comparison to models,
such as those of \citet{ign04}, of clumpy \uchii\ regions.

\subsection{Resolving the Infall Motions} \label{sec:mom1}

\begin{figure}
\epsscale{0.6}
\plotone{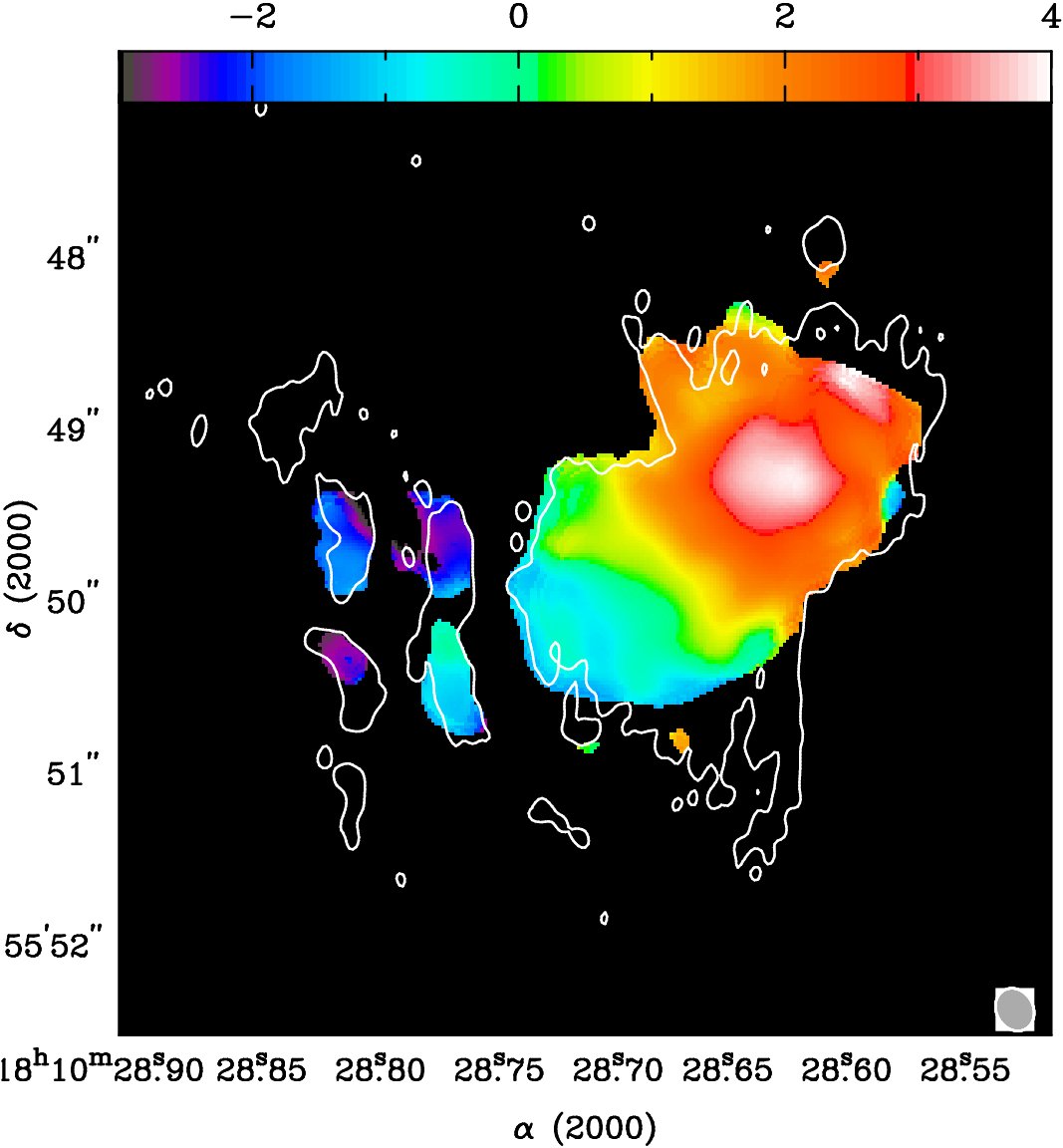}
\caption{The map of the first moment of the main hyperfine component
of the \nhthree\ line. The colors range from \vlsr = -3 to
+4~\kms. The contour, shown for reference, is the 0.8 \mjyperbeam\
contour from the 1.3~cm continuum map shown in Figure
\ref{fig:contmap}. The line data were mapped by the AIPS task IMAGR
with natural weighting and a u-v taper at 750~kilolambda so that the
resulting synthesized beam ($0.''26 \times 0.''22$) better matched the
size scale of velocity pattern.}
\label{fig:mom1}
\end{figure}

Our absorption line data clearly show that infall dominates the
kinematics of the molecular gas. Figure \ref{fig:mom1} is a map of the
first moment (that is the flux-weighted average velocity) of the main
hyperfine component of the \nhthree\ absorption line. While previous
work done at lower angular resolution relied on \pv\ diagrams to
determine the velocity of the gas as a function of position
\citep{ket87b,ket88}, our higher resolution data allow us to map the
velocity of the line at every position in the two dimensional map, not
just along selected one dimensional cuts. The velocity of the
absorbing gas ranges from \vlsr$=+3.5$~\kms\ shown as white in Figure
\ref{fig:mom1}, to \vlsr$=-2.5$~\kms\ shown as purple. The velocity of
the ambient gas, as determined from widespread emission seen at lower
resolution and subsequent radiative transfer modeling is
\vlsr$=-3.0$~\kms\ \citep{ho86,ket90}. Thus the white points in Figure
\ref{fig:mom1}, i.e. the most red-shifted points, show the peak of the
infall. Moving away from that peak position in any direction, the
infall velocity decreases, forming concentric rings of red, orange,
and yellow in Figure \ref{fig:mom1}. The smooth, circular pattern of
radially decreasing red-shift is indicative of spherical infall, in
which the line-of-sight infall velocity gets projected out as one
moves away from the center of infall \citep{ho86}. The velocity
pattern is dominated by this ``bulls-eye'' pattern created by the
spherical infall.

There is also evidence in Figure \ref{fig:mom1} of the rotation that
was seen at larger radii, out to 1~pc \citep{ho86}. Earlier work has
shown that the rotation axis runs from southwest to northeast, with
the rotating gas coming toward the observer in the southeast and going
away from the observer in the northwest \citep{ho86,ho94}. The
rotation is apparent in Figure \ref{fig:mom1} only as a perturbation
to the ``bulls-eye'' pattern of spherical infall. The effect of the
rotation is to shift the apparent center of infall along the plane of
rotation, toward the northwest, where the rotational velocity is away
from the observer along the line of sight. Although the rotation is
evident in Figure \ref{fig:mom1}, the circular velocity at a radius of
0.03~pc is 2~\kms, which is less than the infall velocity. The
molecular gas is therefore spiraling inward to the \uchii\ region
without first settling into a disk.

Despite the rotation seen at larger radii, we see no evidence for the
molecular gas settling into a disk at a radius of 0.03~pc. The optical
depth of the absorbing gas, calculated from the ratio of the
absorption in the main and satellite hyperfine components, shows an
overall radial drop-off from a peak in the center and a slight trend
toward higher optical depths in the plane of rotation, suggesting a
somewhat flattened, but mostly spherical density structure as opposed
to a geometrically thin molecular disk. By contrast, if the molecular
accretion flow were geometrically thin in the plane of rotation of the
core, as in, for instance, a rotationally supported disk, then any
inclination of the molecular disk to the line of sight would result in
high optical depth on one side of the \hii\ region and much lower
optical depth on the other side. This optical depth pattern is not
detected. The more spherical density structure is consistent with the
mainly spherical velocity structure of the flow. While rotation
dominates at larger radii as shown by \citet{ho86, ket88, ho94}, our
data show that at a radius of 0.03~pc, rotation does not dominate. No
disk is formed at that radius. While the rotational velocities
increase inward \citep{ket88}, the infall increases inward more
quickly, and dominates as the gas crosses the \uchii\ region
boundary. The molecular gas apparently spirals into the \uchii\ region
without first settling into a molecular disk. This certainly does not
preclude the possibility of an ionized disk inside the \uchii\ region.

Most of the line absorption must take place very close to the \uchii\
region. Because we see the effects of projection in both the spherical
infall and the rotation components of the velocity structure, the
geometry requires that the absorbing gas be immediately outside the
\hii\ region and wrap closely around the continuum backlight. If the
absorbing gas were not right up against the continuum source, the
radial motions would not show significant projection effects away from
the center of infall, and the velocity would not return to the ambient
velocity at the edge of the absorption, as it does. For this reason,
we can associate the infall and rotation velocities seen in the
absorbing gas with the radius of the \uchii\ region. We also note that
the optical depth falls off by a factor of $\sim 100$ from the center
of the \uchii\ region to the continuum sources A1-A6, which is
consistent with the centrally condensed and centrally heated molecular
core model of \citet{ket90}. A strongly centrally condensed core
naturally locates most of the absorbing material, and therefore most
of the absorption, physically close to the continuum source.

An important connection can be made between the infall we see in the
molecular gas just outside the \uchii\ region, and the motions of
ionized gas inside the \uchii\ region. \citet{ket02a} found that the
\hsixalpha\ radio recombination line showed infall motions in the
ionized gas inside the \uchii\ region. Figure 2 of that paper is a map
of the first moment of the \hsixalpha\ line at 1\arcseconds\
resolution. Comparing \citet{ket02a} and our Figure \ref{fig:mom1}
shows that the \hsixalpha\ and \nhthree\ lines have the same peak
infall velocity at the same position. This shows that the infall in
the molecular gas continues right through the ionization front and
into the ionized gas.

\subsection{The Nature of High Mass Star Formation} \label{sec:nature}

Our maps confirm the presence of simultaneous rotation and infall in
the molecular gas, and ongoing work in the \hsixalpha\ line has shown
both infall over a large area, and outflow in a more localized
jet-like structure in the ionized gas \citep{ket04}. The simultaneity
of these different phenomena with main sequence stars, as evidenced by
the presence of the \uchii\ region, points to a schedule for the
formation of massive stars that is compressed relative to the schedule
for low mass stars. Stellar structure calculations suggest that a
massive star should reach the main sequence before collapse in the
molecular core has finished \citep{nor00,beh01}. We find in our
observations evidence for this compressed evolutionary
sequence. Furthermore, the presence of inward motion in the ionized
gas and its apparent continuity with the molecular accretion flow
suggests that the presence of an \uchii\ region does not necessarily
end accretion as in the classical model for pressure driven expansion
of \hii\ regions. The lack of a molecular disk contrasts with the case
of M17, in which a 20~\msun\ star is accreting through a molecular
disk which extends beyond a radius of 15,000~AU \citep{chi04}. While
the two cores have similar angular velocities, roughly 10~\kmspc, at a
radius of 0.1~pc, the much larger central mass in G10.6 allows infall
to dominate and prevent the formation of a large molecular disk.

The lack of any detectable blue-shifted absorption places a stringent
upper limit on the mass of an expanding molecular shell, which would
be the necessary product of pressure-driven expansion of the \uchii\
region. As noted in Section \ref{sec:intro}, the classical model of an
over-pressured \hii\ region expanding into a molecular medium, results
in a dense, isothermally shocked, molecular layer surrounding the
\hii\ region. This layer should contain nearly all of the molecular
gas originally in the volume now occupied by the \hii\ region. We
estimate that the mass of such a layer, assuming a radius for the
\hii\ region of 0.025~pc, an average molecular gas density over the
volume of the \uchii\ region of $10^6$~\percc, LTE conditions at
150~K, and an \ammonia\ abundance of $10^{-7}$ relative to \htwo\
\citep{van98}, would be 4~\msun resulting in a predicted optical depth
in the main hyperfine component of $\tau \sim 2$. We detect no such
layer in our line data. At the peak of the continuum, our $3\sigma$
upper limit on the optical depth of an expanding molecular layer is
$\tau = 0.05$, corresponding to a shell-mass of roughly 0.1~\msun. We
detect no outward motion; only inward motions in both the molecular
and ionized gas are seen. This stringent upper limit on the mass of
any expanding shell, and the fact that the molecular accretion flow
which we observe at the 0.03~pc scale continues in the ionized gas,
both argue against the pressure-driven expansion model and for the
existence of an ionized accretion flow.

The central mass in this region can be estimated in two ways. First,
we note that the infall peaks around \vlsr$=+3.5$~\kms. The systemic
velocity of the local gas is \vlsr$=-3.0$~\kms\ \citep{ket90}. If the
gas has a free-fall collapse speed of 6.5~\kms\ at a radius of
0.03~pc, the central mass must be about 150~\msun. We can also
calculate the mass of the central star(s) by inferring the rate of
Lyman continuum photon production. Because the ionized gas is
infalling, we assume the density goes like $\rm{n(r) =
n_0(\frac{r_0}{r})^{(3/2)}}$ with $\rm{r_0 = 0.03}$~pc, and $n_0 =
10^5$~\percc\ .  Then, following \citet{ket03}, we balance ionizations
with recombinations as
\begin{displaymath}
\rm{N_{Ly} = \int_{R_*}^{R_{HII}} \alpha 4 \pi r^2 n^2(r) dr = 4 \pi r_0^3 n_0^2 \alpha~ln(\frac{R_{HII}}{R_*})} \nonumber
\end{displaymath}
where $\rm{N_{Ly}}$ is the number of Lyman continuum photons,
$\rm{R_*}$ is the radius of the star, $\rm{R_{HII}}$ is the radius of
the \hii\ region, and $\alpha$ is the recombination
coefficient. Assuming $\rm{R_*= 10 R_{\odot}}$, $\rm{R_{HII}} =
0.03$~pc as above, and $\rm{\alpha= 2 \times 10^{-13}~cm^3~s^{-1}}$,
we get $\rm{N_i = 2.2 \times 10^{50} s^{-1}}$. This corresponds to 4
O4V stars with a total mass of 175 \msun, and a total luminosity of
$4.4 \times 10^6$~\lsun\ \citep{vac96}. This combination of stars,
which matches the Lyman continuum output and, roughly, the central
mass as derived from the free-fall velocity, is not unique. However,
we can be certain that at least several stars must be responsible for
such a large Lyman continuum emission rate.

The complicated structure of the ionized gas suggests clumpy structure
in the surrounding molecular gas. This clumpy structure at very high
density ($> 10^6$~\percc) is consistent with the idea that this single
HMC may be capable of forming many stars. The central \uchii\ region
has formed at the center of the HMC, but must contain more than one
star. The overall contraction of the core has already formed multiple
stars, and may fragment further at smaller scales.

\bibliographystyle{apj}
\bibliography{bib_entries}

\end{document}